\documentclass{elsart}

\usepackage{epsfig}
\usepackage{amssymb, amsmath}
\begin{document}
\begin{frontmatter}
\title{Evidence for the positive-strangeness 
pentaquark \boldmath$\Theta^+$\unboldmath\ in photoproduction
with the SAPHIR detector at ELSA\thanksref{DFG}}
\collab{The SAPHIR Collaboration}
\author[Bonn1]{J.\,Barth},
\author[Bonn1]{W.\,Braun\thanksref{left}},
\author[Bonn2]{J.\,Ernst},
\author[Bonn1]{K.-H.\,Glander},
\author[Bonn1]{J.\,Hannappel\thanksref{left}},
\author[Bonn1]{N.\,J{\"o}pen},
\author[Bonn2]{H.\,Kalinowsky},
\author[Bonn1]{F.\,Klein},
\author[Bonn2]{E.\,Klempt},
\author[Bonn1]{R.\,Lawall},
\author[Bonn2]{J.\,Link\thanksref{left}},
\author[Bonn1]{D.\,Menze},
\author[Bonn1]{W.\,Neuerburg\thanksref{left}},
\author[Bonn1]{M.\,Ostrick},
\author[Bonn1]{E.\,Paul},
\author[Bonn2]{H.\,van\,Pee},
\author[Bonn1]{I.\,Schulday},
\author[Bonn1]{W.\,J.\,Schwille},
\author[Bonn1]{B.\,Wiegers\thanksref{left}},
\author[Bonn1]{F.\,W.\,Wieland},
\author[Bonn1]{J.\,Wi\ss{}kirchen\thanksref{left}},
\author[Bonn1]{C.\,Wu\thanksref{left}}.

\address[Bonn1]{Physikalisches Institut, Bonn University, Bonn, Germany}
\address[Bonn2]{Helmholtz-Institut f{\"u}r Strahlen- und Kernphysik, 
Bonn University, Bonn, Germany}
\thanks[DFG]{Supported by Deutsche Forschungsgemeinschaft within
the Schwerpunktprogramm SPP 1034 KL 980/2-3}
\thanks[left]{No longer working at this experiment}
\date{\today}
\maketitle
\vskip 5mm
\begin{abstract}
The positive--strangeness baryon resonance $\Theta^+$ is observed
in photoproduction of the $\rm nK^+K^0_s$ final state with the 
SAPHIR detector at the Bonn ELectron Stretcher Accelerator ELSA.
It is seen as a peak in the $\rm nK^+$ invariant mass distribution
with a $4.8\sigma$ confidence level. We find a mass 
$\rm M_{\Theta^+} = 1540\pm 4\pm 2$\,MeV and an upper
limit of the width $\rm \Gamma_{\Theta^+} < 
25$\,MeV  at 90\%  c.l. From the absence
of a signal in the $\rm pK^+$ invariant mass distribution in 
$\rm\gamma p\to pK^+K^-$ at the expected strength
we conclude that the $\Theta^+$ 
must be isoscalar.\\
\vspace{.5cm}
\end{abstract}

\end{frontmatter}

\section{Introduction}
The quark model has been extremely successful in accounting for
the observed meson and baryon states and their quantum numbers.
Ground--state baryons are grouped into an octet {\bf 8} and a 
decuplet {\bf 10} with total angular momentum $\rm J^P=1/2^+$ 
and $3/2^+$, respectively. The successful prediction of the 
$\Omega^-$ and of its mass \cite{Gell-Mann:nj} and its subsequent 
experimental discovery \cite{Barnes:pd} was a cornerstone
in establishing SU(3) symmetry. The pattern of excited baryons 
was understood in terms of three quarks obeying the Pauli 
principle (see, e.g. \cite{Isgur:1977ef} and refs. therein).
The discovery of the J/$\psi$ \cite{Aubert:1974js,Augustin:1974xw}
and $\Upsilon$ \cite{Herb:ek} families gave final support to the quark 
model. The development of QCD as renormalizable field theory 
\cite{Weinberg:1973un,Fritzsch:1973pi} and the observation of 
the partonic structure of nucleons in deep inelastic scattering 
\cite{bjorken} seemed to pave the way for an eventual understanding 
of the structure of hadronic matter.
\par
A completely different approach to understand the nucleon was 
proposed by Skyrme\cite{Skyrme:vq,Skyrme:vh}. He suggested 
that the low-energy behaviour of nucleons can be viewed as 
spherically symmetric soliton solution of the pion field. 
In this view, the baryonic nature of the nucleon is not 
due to quarks carrying baryon quantum number $B=1/3$. 
Instead, the baryon number is interpreted as topological 
quantum number of the pion field \cite{Witten:tw}.
\par
A fascinating consequence of this view 
is the prediction of an anti--decuplet $\bf\overline{10}$. 
It contains a state with exotic quantum numbers, 
with strangeness $S=+1$ and $\rm J^P=1/2^+$ \cite{Chemtob:ar}. 
Walliser estimated the mass of this exotic state to be at about
1700\,MeV \cite{Walliser:vx}. Diakonov {\it et al.} assigned 
the $\rm N(1710)\,P_{11}$ to the non-strange
member of the anti--decuplet thus fixing
the mass of the exotic $\Theta^+$ to 1530\,MeV 
\cite{Diakonov:1997mm}, a value consistent with the prediction
of Weigel one year later \cite{Weigel:1998vt}. Due to its 
intrinsic strangeness and low mass, the only allowed strong--interaction
decay modes of the $\Theta^+$ are $\rm nK^+$ and $\rm pK^0$. Its
width is predicted to be narrow. In quark models, the $\Theta^+$
has a minimal quark content $\rm uudd\bar s$ and is called 
pentaquark. In the soliton model, it 
is the lowest--mass state beyond the conventional octet 
and decuplet baryons. A recent survey can be found in 
\cite{Walliser:2003dy}. 
\par
These predictions stimulated an active search for the $\Theta^+$.
At SPring-8(LEPS),  a $4.6\sigma$ $\rm nK^+$ peak was reported in
the reaction $\rm \gamma ^{12}C \to K^+K^- X$ \cite{Nakano:2003qx}. 
Cuts were applied to select quasi--free processes on the neutron.
After correction for the Fermi momentum, 
the mass was determined to  $(1540\pm 10)$\,MeV.
The observed width was compatible with the instrumental resolution. 
An upper limit of $\Gamma <25$\,MeV was derived. 
The DIANA collaboration at ITEP \cite{Barmin:2003vv}
reported a $\rm K^0_sp$ enhancement produced in low-energy $\rm K^+$\,Xe
collisions, with $\rm M = (1539\pm 2), \Gamma < 9$\,MeV and a 
statistical significance of $4.4\sigma$. 
\par
At Jlab, a significant $\Theta^+$ peak was observed with the CLAS detector
in the exclusive reaction $\rm d(\gamma, K^+K^-p)n$. The $\Theta^+$
is photoproduced off the neutron and subsequently decays into $\rm nK^+$.
Only in case of $\rm K^-p$ rescattering, the proton received 
sufficient momentum to be detected; the neutron was identified by 
the missing mass \cite{Stepanyan:2003qr}. 
The peak in the $\rm nK^+$ mass distribution 
had a statistical significance of $5.3\pm 0.5\sigma$; it was 
found at a mass of $(1542\pm 5)$\,MeV, its width was compatible 
with the resolution (FWHM = 21\,MeV). 
The authors did not see a peak in the $\rm pK^+$ mass distribution.
Since the CLAS acceptance for $\rm pK^+$ is not 
the same as that for $\rm nK^+$,   
they cautiously suggested the $\Theta^+$ should have isospin zero. 
A preliminary study of the $\rm n\pi^+K^-K^+$ final state 
produced by photoproduction off protons by
the CLAS collaboration \cite{Clas-prop} suggests a $4.3\sigma$ peak in 
the $\rm nK^+$ invariant mass distribution recoiling (in the centre--of--mass
system) against a $\rm K^*$. 
\par
\begin{figure}[h!t]
\begin{center}
\epsfig{file=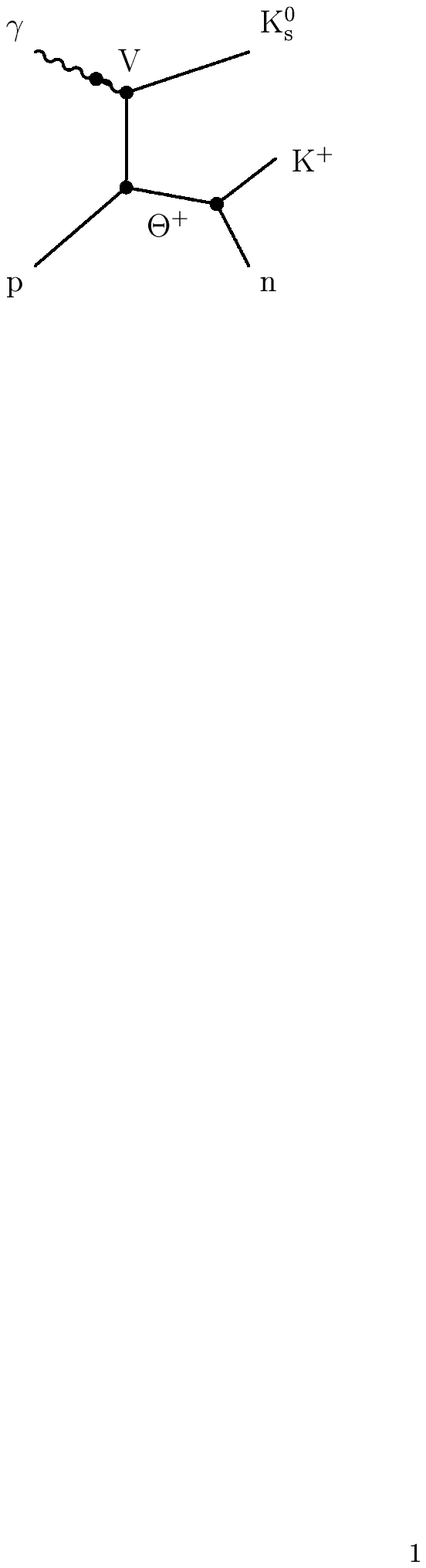,width=0.4\textwidth,clip=}
\end{center}
\label{diag1}
\caption{\it Diagram through which the $\Theta^+$
could be produced. The V represents a vector meson.
}
\end{figure}

In this letter we present evidence for
the $\Theta^+$ in photoproduction of 
the $\rm nK^+K^0_s$ final state off protons. 
A diagram which may contribute is shown in Fig. \ref{diag1}.

\section{Experimental method}
The reaction
\begin{equation}
\rm\gamma p \to nK^0_sK^+ \quad {\mathrm \ with \ the \ decay} \quad
K^0_s\to\pi^+\pi^-
\end{equation}
was measured with the SAPHIR detector \cite{Schwille:vg} at ELSA.
SAPHIR is a magnetic spectrometer covering a solid angle of 
approximately  $0.6\times4\pi$\, and the full polar range
$0<\vartheta <\pi$. 
The ELSA electron beam produced photons via
bremsstrahlung in a copper foil radiator. The energies of the
scattered electrons were determined in the tagging system
for photon energies from 31\,\% to 94\,\% of the incident electron
energy which was  2.8 GeV for the data presented in this letter.  
The photon beam passed through a liquid hydrogen target (3\,cm in
diameter and 8\,cm in length). Non--interacting photons
were detected in a photon counter. The coincidences of tagger 
and photon counter determined the photon flux.  
\par
The target was located in the centre of a central drift chamber with 14 
cylindrical layers. Charged particles
leaving the target were bent in the field of a C-shaped magnet allowing for
a measurement of momentum and sign of charge.  The momentum 
resolution $\Delta p/p$ for 300\,MeV/c particles in the central drift
chamber was approximately 2.5\,\%. In forward direction a planar drift
chamber improved  the momentum resolution to about $1$\,\%. The
central drift chamber was surrounded by a scintillator wall to
measure the time of flight (ToF). The minimum momentum
to detect tracks of charged pions and protons was 50 and 150 MeV/c,
respectively, with  a track detection threshold depending
on the position of the production vertex in the target. 
The data presented here were taken in 1997 and 1998 with a trigger 
requiring at least two charged particles. About $1.33\cdot 10^8$ events
were recorded. The data were used to determine cross sections
for photoproduction of strange mesons and baryons 
\cite{Bennhold:1997mg},\cite{Tran:qw},\cite{Goers:sw}, and 
of $\omega$ \cite{Barth:?}, $\eta'$ \cite{Plotzke:ua}, and 
$\Phi$ \cite{Barth:bq} mesons.

\section{Evidence for the $\Theta^{+}$}

We observe the $\Theta^+$ in the reaction 
\begin{equation}
\rm \gamma p \to \Theta^{+} K^0_s ; \qquad\ \Theta^{+}\to nK^+ ;
\qquad K^0_s\to\pi^+\pi^-
\label{eq:plus}
\end{equation}
The neutron is identified in a kinematical fit using one
constraint; the neutron momentum is reconstructed
from energy and momentum conservation. 
Fig. 2a shows the $\rm nK^+$ invariant mass distribution
after cuts explained below. A clear signal is observed
at $\sim 1540$\,MeV above a smooth background. The fit ascribes
$63\pm 13$ events ($4.8\sigma$) to the peak. By counting the four 
bins around the $\Theta^+$ centre
we get 55 events above background
and 56 below, corresponding to a statistical significance
of  $5.2\sigma$.
\par
\begin{figure}
\begin{tabular}{cc}
\includegraphics[width=0.98\textwidth]{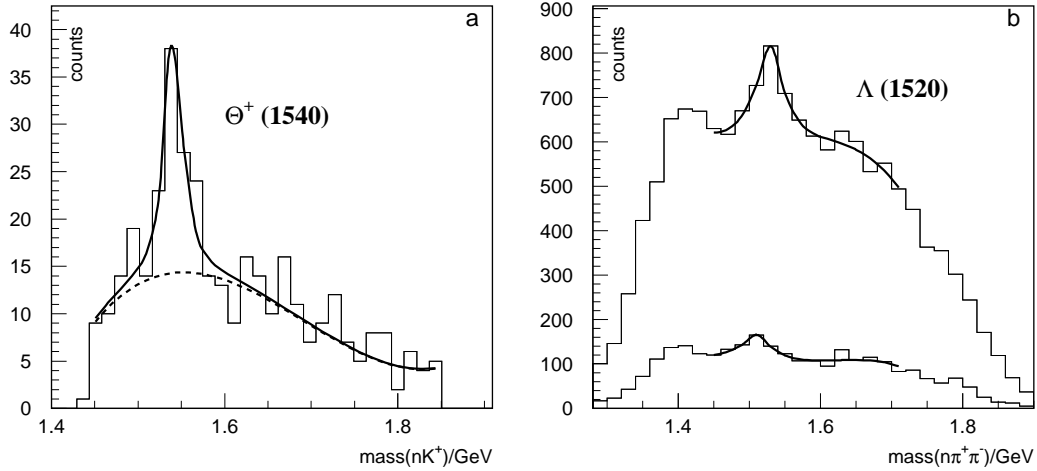}
\end{tabular}
\label{pic:plus}
\caption{\it a) The $\rm nK^+$ mass distribution 
after cuts in the $\pi^+\pi^-$ mass distribution
and in the $\rm K^0_s$ production angle.
b) The $\rm n\pi^+\pi^-$ invariant mass distribution 
showing the $\Lambda (1520)$ without (upper curve) and with 
(lower curve) $\cos\vartheta_{\pi^+\pi^-}$ cut.
The solid lines represent fits using a Breit--Wigner distribution 
(convoluted with a resolution function) plus polynomial background.
}
\end{figure}
To determine mass and width we simulated reaction (\ref{eq:plus})
by Monte Carlo, generating a narrow $\Theta^+$. The resulting 
$\rm nK^+$ mass distribution (after event reconstruction) is used 
as resolution ($\rm FWHM = 23$\,MeV) function which is folded 
with a Breit--Wigner 
distribution to fit the data. We find
\begin{equation}
\rm M_{\Theta^+} = 1540\pm 4\pm 2\,MeV \qquad\ 
\rm \Gamma_{\Theta^+} < 25\,MeV \ at\ 90\% \ c.l.
\label{mass}
\end{equation}
The errors quoted are the statistical and systematic uncertainties.
The statistical error includes variation of the results 
when fit range or background function are changed. 
The systematic error is estimated from the comparison of
SAPHIR and PDG masses for known stable particles. 
\par
A series of cuts has been applied to arrive at Fig. 2a.
Events are subjected to kinematical fits to the hypotheses 
$\rm \gamma p\to n\pi^+\pi^-K^+$ and 
$\rm \gamma p\to n\pi^+\pi^-\pi^+$. The
confidence level for the former hypothesis is required to
exceed 1\%, while the latter hypothesis should have
a probability of less than 1\%. To further clean the data, we
rejected events which passed one of the following
hypotheses (*):  
$\rm \gamma p\to p\pi^+\pi^-$, $\rm \gamma p\to pK^+K^-$,
$\rm \gamma p\to pe^+e^-$, $\rm \gamma p\to p\pi^+\pi^-\pi^0$, 
$\rm \gamma p\to p\pi^-K^+$, $\rm \gamma p\to p\pi^-\pi^0K^+$.
\par
We require identification of two pions and of one 
$\rm K^+$ by time--of--flight. A cut in the $\pi^+\pi^-$
invariant mass from 480 to 518\,MeV has been applied which
enhances the fraction of  $\rm nK^+K^0_s$ events.
We observe that the $\rm K^0_s$ from $\Theta^+$ production
is preferentially produced in forward direction. 
A $\cos\vartheta_{\rm K^0_s} >0.5$ cut reduces the signal strength by
a factor of 2 only while the background is reduced to 1/5.
$\vartheta_{\rm K^0_s}$ denotes the center--of--mass angle
between the $\rm K^0_s$ and the photon beam direction. 
In the $\rm n\pi^+\pi^-$ mass distribution (Fig. 2b), 
the $\Lambda(1520)$ is clearly seen with $530\pm 90$ events. 
The $\cos\vartheta_{\pi^+\pi^-}>0.5$ cut considerably reduces 
the $\Lambda(1520)$ signal.  
\par
The size of the $\rm nK^+$ peak in Fig. 2a does not depend
on the kinematical fits and is not generated by them.
The signal can also be seen without the kinematical anti--cuts 
(*) and without the cut in $\cos\vartheta_{\rm K^0_s}$, 
yet above a larger background. All kinematical
fits reduce the background but do not lead to noticeable
losses in the signal strength; the $\cos\vartheta_{\rm K^0_s}$
cut reduces the signal by about a factor 2.
\par
Fig. 3a presents the $\pi^+\pi^-$ mass distribution 
after applying a mass cut in the $\rm nK^+$ 
invariant mass from 1.515 to 1.559\,MeV. A clear $\rm K^0_s$ signal 
is observed. We now verify that the peaks in Fig. 2a 
and 3a are correlated. 
Fig. 3b shows the side--bin subtracted $\rm nK^+$ mass distribution.  
The $\rm K^0_s$  is defined by the 480 to 516\,MeV mass interval,
the two side bins, counted with half weight, extend from
417 to 453 and from 543 to 579\,MeV, respectively. In the subtracted
spectrum (Fig. 3b) the $\Theta^+$ is seen above a rather small
background. The background resembles
the distribution expected from the reflection of the
$\rm\Lambda (1520)K^+$;
its content agrees with 
the number of $\rm\Lambda (1520)$ events in the lower curve of 
Fig. 2b. We conclude 
that the $\rm nK^+K^0_s$ final state is reached dominantly
via the intermediate states $\rm\Theta^+K^0_s$ or 
$\rm\Lambda (1520)K^+$. Note that fig. 3b was not used to determine 
the signal strength.
\begin{figure}
\begin{tabular}{cc}
\includegraphics[width=0.98\textwidth]{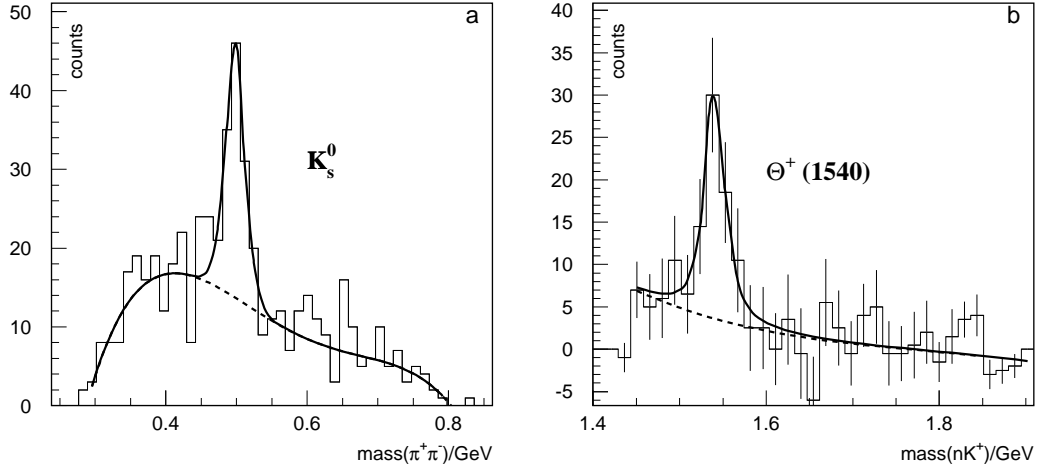}
\end{tabular}
\label{pic:plus2}
\caption{\it a) The $\pi^+\pi^-$ mass distribution after a cut in the
$\rm nK^+$ mass distribution selecting $\Theta^+$ events
with a fit using a resolution function and a polynomial background.
b) The $\rm nK^+$ mass distribution 
after side-bin subtraction of the background under the $\rm K^0_s$.
The residual background is the reflection from the $\Lambda (1520)$
surviving the $\cos\vartheta_{\rm K^0_s}$ cut.
The solid line represents a fit using a Breit--Wigner distribution 
(convoluted with a resolution function) plus polynomial background.
 }
\end{figure}
\par
From the number of $\Theta^+$ events observed in the full angular range, the mean 
cross section for $\Theta^+$ production in the photon energy range
from threshold (1.74\,GeV) to 2.6\,GeV is estimated to about
$200$\,nb, apparently rising with energy. 
It may be compared to those for other reactions with
strange particles in the final state like $\rm\gamma p\to K^+\Lambda$, 
$\rm\gamma p\to K^+\Sigma$, and $\rm\gamma p\to K^+\Lambda (1520)$
which are of the order of 800 -- 1200\,nb at $\rm E_{\gamma}=2$\,GeV. 
\section{The isospin of the $\Theta^+$}

If the $\Theta^+$ is a member of the anti--decuplet, it has 
quantum numbers $\rm \Theta^+(1540)P_{01}$.
In particular it is predicted to be an isoscalar. 
A conventional explanation would interpret the $\Theta^+$
as NK bound state. Isospins 0 and 1 are
both possible; isospin 1 would lead to three
charge states $\Theta^0,\Theta^+,\Theta^{++}$.
Capstick, Page and Roberts \cite{Capstick:2003iq}
point out that the $\Theta^+$ could be a member of an isospin
quintet with charges from -1 to +3 where the $Q=+3$ state 
has a $\rm (uuuu\bar s)$ quark--model configuration.  
An isotensor resonance could be produced with an isospin
conserving amplitude as depicted in Fig.~4 (left). 
Decays of an isotensor $\Theta^{+}$ into $\rm pK^0$
and $\rm pK^+$  are isospin violating; hence an isotensor $\Theta^{+}$ 
is expected to be narrow \cite{Capstick:2003iq}. 
\par
The photoproduction of the $\rm pK^+K^-$ final state has   
been studied recently by us \cite{Barth:bq}, with the 
focus on $\Phi$ production. We use these data to search for
the doubly charged $\Theta^{++}$ in the reaction chain
\begin{equation}
\rm \gamma p \to \Theta^{++} K^- ; \qquad\ \Theta^{++}\to pK^+.
\label{eq:plusplus}
\end{equation}
In analogy to the $\rm K^0_s$ above, we have applied a cut  
$\cos\vartheta_{\rm K^-}>0.5$ to select forward ${\rm K^-}$. 
Fig.~4 (right) shows the resulting $\rm pK^+$
invariant mass spectrum. The data are consistent with a small 
structure at 1540\,MeV. The fit assigns $75\pm 35$ 
events to it. 
\begin{figure}[h!t]
\begin{tabular}{cc}
\epsfig{file=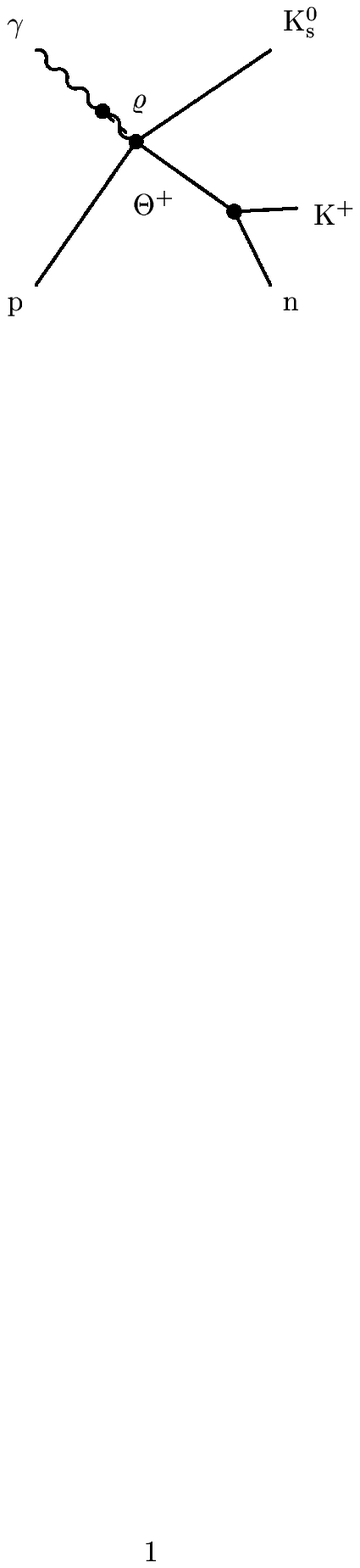,width=0.4\textwidth,clip=}&
\includegraphics[width=0.5\textwidth]{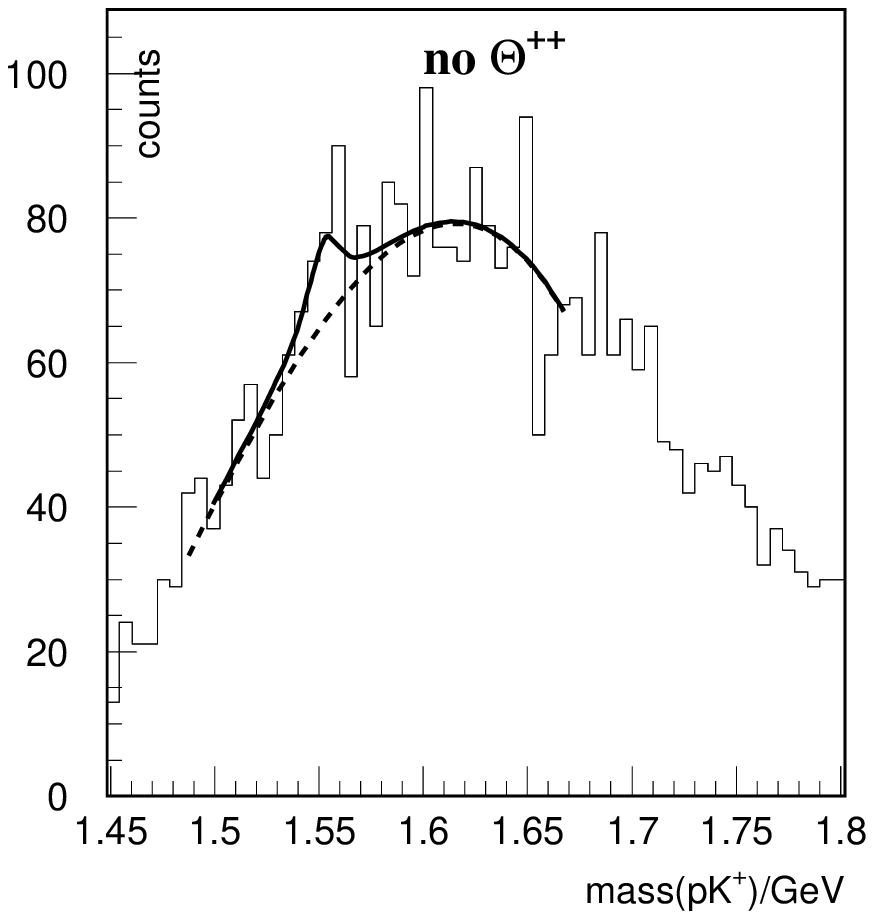}
\end{tabular}
\label{pic:plusplus}
\caption{\it Left: Diagram allowing for production of an 
isotensor resonance. \qquad\quad
Right: The $\rm pK^+$ invariant mass distribution in the
$\rm pK^+K^-$ final state. The solid line represents a fit using 
a Breit--Wigner distribution (convoluted with a resolution function) 
plus polynomial background. If the $\Theta^+$ would be 
an isovector or isotensor state, it would yield a much larger
peak containing more than 5000 events.
}
\end{figure}
\par
The number of $\Theta^{++}$ events we expect depends on the 
isospin: 
Clebsch--Gordan coefficients favour $\Theta^{++}$ over $\Theta^{+}$
by a factor 3 (for I=2) or 4 (for I=1) while the  $\Theta^{++}$
does not exist in case of I=0. Experimentally, the final
state $\rm K^+K^-$ offers an additional factor 3 since 1/2 of
all $\rm K^0$ are produced as $\rm K^0_L$ and escape undetected,
and the $\rm K^0_s$ is observed only with a fraction $\sim 2/3$.
In case the $\Theta^{+}$ is an isovector or isotensor we should observe
9 or 12 times more $\Theta^{++}$ events, respectively. Also, the
SAPHIR acceptance is considerably larger for the fully constrained
$\rm pK^+K^-$ events. We thus expect a peak with
more than 5000 $\Theta^{++}$ in Fig. 4 (right), 
far above the observed level. 
We conclude that the $\Theta^{+}$ is isoscalar.

\section{Summary and conclusions}

We have studied photoproduction of the $\rm nK^+K^0_s$ final state.
In the $\rm nK^+$ invariant mass distribution we observe a $4.8\sigma$
peak at a mass of $\rm M_{\Theta^+} = 1540\pm 4\pm 2$\,MeV. 
The peak is strikingly narrow:  we quote an upper limit 
(at 90\%  c.l.) of $\rm \Gamma_{\Theta^+} < 25$\,MeV. Mass and width are
compatible with results obtained elsewhere. Here, the $\Theta^+$
is observed for the first time in the reaction $\rm\gamma p\to nK^+K^0_s$.
The cross section for $\rm\gamma p\to\Theta^+\overline K^0$ in the photon
energy range from 1.7 to 2.6\,GeV is of similar size as the cross
section for $\phi$ photoproduction and 
smaller by a factor 4 than for other final states with open strangeness like
$\rm\gamma p\to\Lambda K^+$, $\rm\gamma p\to\Sigma K^+$, or
$\rm\gamma p\to\Lambda(1520) K^+$.
\par
The $\Theta^+$ has strangeness $S=+1$ and cannot be a three-quark
baryon. In the quark picture, it is a pentaquark $\rm (uudd\bar s)$.
In the related reaction $\rm\gamma p\to pK^+K^-$, we have searched for
a doubly charged isospin partner of the $\Theta^+$. From the absence
of a signal at the expected strength we deduce that the $\Theta^+$
is an isoscalar.

\section*{Acknowledgements}
We would like to thank the technical staff of the ELSA machine group
for their invaluable contributions to the experiment. We gratefully 
acknowledge the support by the Deutsche Forschungsgemeinschaft
in the framework of the Schwerpunktprogramm "Investigations of the hadronic
structure of nucleons and nuclei with electromagnetic probes"
(SPP 1034 KL 980/2-3).

\end{document}